\documentclass[11pt]{article} 
\usepackage{amsfonts,amscd,amsmath,graphicx}

\def\NP#1#2{ Nucl.Phys. B#1 (#2)} 
\def\PL#1#2{ Phys.Lett. B#1 (#2)}

\def\PR#1#2{Phys.Rev. D#1 (#2)} 
\def\IJMP#1#2{ Int.J.Mod.Phys. A#1 (#2)}

\def\HP#1#2{ JHEP #1 (#2)} 
\def\ap{ \alpha^{\prime}} 

\def\pd{\partial}

\newcommand{\ep}{\text e}
\newcommand{\oh}{\frac{1}{2}}

\def\3{\Phi^3}

\title{A Note on Nonperturbative Instability in String Theory}
\author{Oleg Andreev\thanks{e-mail:  andreev@physik.hu-berlin.de}
\thanks{Permanent address: Landau Institute, Moscow, Russia}
\\ \\
Humboldt--Universit\"at zu Berlin, Institut f\"ur Physik\\
Invalidenstra\ss e 110, D-10115 Berlin, Germany}

\date{}
\begin{document} 

\vspace{-8cm} 
\maketitle 
\begin{abstract} 
We investigate the possibility that stringy nonperturbative instabilities are described by worldsheet methods. We 
focus on the case of open bosonic string theory, where the D-instanton plays a role of the bounce, i.e. it describes 
barrier penetration. In the process, we compute the exponential factor in a decay probability. 
\end{abstract}

\vspace{-10cm}
\begin{flushright}
hep-th/0112088      \\
HU Berlin-EP-01/59
\end{flushright}
\vspace{9cm}

It is well-known since the seventies that the  perturbative string S-matrix defined as expectation 
values of worldsheet vertex operators reproduces field theory scattering amplitudes \cite{scherk}. If one thinks of 
string theory as field theory with infinite number of fields living in higher dimensions then this statement is equivalent to 
saying that scattering amplitudes of string theory near its trivial perturbative vacuum are reproduced by 
worldsheet methods. It is a big problem to understand the vacuum structure of string theory. It is clear that some vacua 
may be unstable in perturbative or even in nonperturbative sense. In the last case, it seems natural to ask whether 
one can compute decay probabilities as for example it was done by instanton methods in ordinary field 
theory \cite{okun}. It is clear that this is in principle a solvable problem. However, in practice it turns out to be very 
hard to deal even with several fields not saying an infinite number of them. A hope maybe that worldsheet methods 
are appropriate again. A good motivation for this comes from the fact that there are representations of string 
theory effective actions $S$ via worldsheet objects \cite{Z,w}. These representations are based on 
partition functions of strings propagating in background fields. Moreover, the actions evaluated at 
solutions of the corresponding equations of motion coincide with the partition functions namely, 

\begin{equation}\label{s=pf}
S(\lambda^i_\ast)=c Z(\lambda^i_\ast)
\quad,
\end{equation}

\noindent where $\lambda^i$ are string fields and $Z$ is the partition function. Here, we also include a normalization 
constant $c$.  The extrema of the actions have the known meaning within the worldsheet theory: they are 
conformal backgrounds. So, the right hand side of Eq.\eqref{s=pf} represents a partition function 
of two-dimensional theory at its fixed point. This simplifies explicit computations of partition functions since 
there are no UV divergencies anymore. Let us now assume that equations of motion associated with the 
Euclidean action $S_{\text{\tiny E}}$ obtained from $S$ by analytic continuation admit a solution 
$\lambda^i_{\text{\tiny B}}$ which from the field theory point of view can be recognized as 
the bounce, i.e. it describes a decay of some false vacuum $\lambda_0^i$ \cite{okun}. Then, 
assuming that the relation \eqref{s=pf} is also valid in the Euclidean case, we can immediately get the following 
representation for the exponential factor in a decay probability per unit time per unit volume

\begin{equation}\label{dp}
w\sim\exp\bigl[-cZ_{\text{\tiny E}}(\lambda^i_{\text{\tiny B}})\bigr]
\quad.
\end{equation}
Notice that we do not include $S_{\text{\tiny E}}(\lambda^i_0)$ into the exponent.

The purpose of this note is to give an example of explicit computations.  To do so, we consider open bosonic string theory 
where a big progress has been recently achieved in understanding of D-brane decay as open string 
tachyon condensation \cite{as}. In particular, this was achieved by using a toy field 
theory model \cite{z} and a background independent open string field theory \cite{w}. Indeed, they turned out to be
 useful tools to gain intuition on the physics and carry out some explicit calculations. It turns out that they are 
useful for our purpose as well. 

To gain some intuition, we begin with a scalar field theory in Euclidean $p+1$ dimensional space with action 

\begin{equation}\label{tm}
S_{ft}=\tau_{p}\int d^{\,p+1}x\,\ep^{-T}\Bigl(1+T+\oh\ap\pd_iT\pd_i T\Bigr)
\quad.
\end{equation}

\noindent This theory was used as a toy model for tachyon condensation on unstable branes in \cite{z}. In this context, 
it describes the open string tachyon living on an unstable $p$-brane whose tension is $\tau_p$. Note that the 
tension includes a factor of the dilaton (string coupling constant $g$).

One remarkable observation is that the theory belongs to a set of field theory models whose equations of motions 
admit exact solutions \cite{gold}. In particular, a set of exact spherically symmetric solutions associated with the action 
\eqref{tm} is given by 

\begin{equation}\label{sol}
T_n(x)=\frac{1}{2\ap}\Bigl(x_0^2+\dots+x_{n-1}^2\Bigr)-n
\quad,
\end{equation}

\noindent where $n$ ranges from $1$ to $p+1$. In the context of tachyon condensation, these solutions are 
interpreted as the lower dimensional branes. Indeed, they almost reproduce the famous descent relations for the 
D-brane tensions \cite{z}. 

It is easy to find the bounce among the set of the solutions. It corresponds to  $n=p+1$. Indeed, only in this case 
the Euclidean action \eqref{tm} evaluated at the solution is finite. Moreover, one can immediately check that 
$T_{p+1}$ obeys the boundary conditions for the bounce in the sense of Coleman \footnote{To complete the picture, one 
can compute the spectrum fluctuations near $T_{p+1}(x)$ and show that it includes only one negative 
mode as it should be for the bounce. In doing so, there is a great simplification as the corresponding equation reduces 
to the equation for a harmonic oscillator  (see, e.g., \cite{z}).}

\begin{equation}\label{bounce}
\lim_{x_0\rightarrow\pm\infty}T_{p+1}(x)=+\infty
\quad,\quad
\pd_0 T_{p+1}\big\vert_{x_0=0}=0
\quad,
\end{equation}

\noindent where $\pd_0=\pd/\pd x_0$ and $x_0$ is treated as Euclidean time. 

At this point we should mention that in the context of tachyon condensation $T_{p+1}(x)$ is called as the D-instanton. 
Thus, what we have learned from this toy field theory model is a hint on the physical meaning of the D-instanton: 
it might describe a decay of an unstable vacuum through barrier penetration. In our case, the unstable vacuum 
corresponds to $T=+\infty$ (see Fig.1).
%
\begin{figure}[ht]
\begin{center}
 \includegraphics{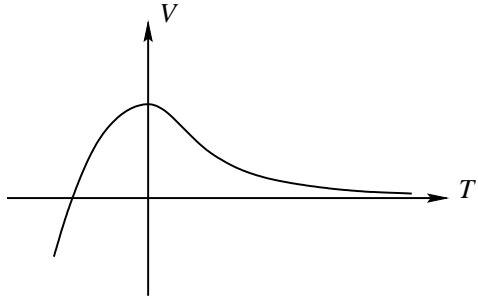}
 \caption{The nonderivative part of the Lagrangian, $V$, for a theory with the Euclidean action \eqref{tm}. 
It gives the tachyon potential.}
 \end{center}
\end{figure}

To complete our discussion of the field theory model, let us compute the exponential factor in a decay probability of 
the unstable vacuum. Evaluating the action at $T_{p+1}(x)$, we obtain 

\begin{equation}\label{d-ins}
S_{ft}(T_{p+1})= \tau_{p} \,\ep^{p+1} \bigl(2\pi\ap\bigr)^{\frac{p+1}{2}}
\quad
\end{equation}

\noindent that results in the following expression for the exponential factor

\begin{equation}\label{prob}
w\sim\exp{\Bigl[-\tau_{p} \,\ep^{p+1} \bigl(2\pi\ap\bigr)^{\frac{p+1}{2}}\Bigr]}
\quad.
\end{equation}
Since $S_{ft}$ evaluated at $T=+\infty$ vanishes, there is no the corresponding contribution in \eqref{prob}.

So far we have just noticed that the D-instanton might be interpreted as a bounce in the sense that 
it describes a decay of some unstable vacuum. However, our discussion given within the field theory model has two 
disadvantages: it can not in principle provide us with the desired representation of the decay factor \eqref{dp}. 
The relation of the solution $T_{p+1}$ with the D-instanton might seem not sufficiently  convincing. Fortunately, 
both of these disadvantages disappear in the background independent open string field theory \cite{w}. To the leading 
order in derivatives, its Euclidean action is simply \cite{gera,kud}

\begin{equation}\label{ws}
S_{\text{\tiny E}}=\tau_{p}\int d^{\,p+1}x\,\ep^{-T}\Bigl(1+T+\ap\pd_iT\pd_i T+\dots\Bigr)
\quad,
\end{equation}

\noindent where the dots stand for an infinite number of higher derivative terms.  These terms  can in principle 
be considered as a result of integration over the other open string modes that modifies the action \eqref{tm}. It turns 
out that this modification of the action does not have a strong influence on the existence of a set of exact spherically 
symmetric solutions like \eqref{sol}. We now have

\begin{equation}\label{sol1}
T_n(x)=\frac{t}{2\ap}\Bigl(x_0^2+\dots+x_{n-1}^2\Bigr)+a
\quad,
\end{equation}

\noindent where $a$ and $t$ are some parameters which will be determined later.

We will again specialize to $T_{p+1}$ because it results in a finite action and obeys the boundary conditions 
\eqref{bounce}. Due to these reasons, we will call it the bounce. It turns out that the action evaluated at the bounce can be 
rewritten as a function of the parameters in the following form \cite{w}

\begin{equation}\label{w}
S_{\text{\tiny E}}(a,t)=\tau_{p}\Bigl[1+\beta_a\frac{\pd}{\pd a}+\beta_t\frac{\pd}{\pd t}\Bigr]Z_{\text{\tiny E}}(a,t)
\quad,
\end{equation}

\noindent where $\beta_a=-a-(p+1)t,\,\,\beta_t=-t$ and 
$Z_{\text{\tiny E}}(a,t)=\bigl(2\pi\ap/t\bigr)^{\frac{p+1}{2}}\ep^{-a}\bigl[\ep^{\gamma t}\,\Gamma(1+t)\bigr]^{p+1}$. 
$\gamma$ denotes the Euler's constant. 

The parameters in \eqref{w} are determined by demanding that $S_{\text{\tiny E}}(a,T)$ is stationary under 
their variations. It is a simple task to do so for $a$ \cite{kud}. Indeed, in this case a simple algebra leads 
to $a(t)=(p+1)\Bigl[-t+t\frac{\pd}{\pd t}\bigl(\gamma t+\ln\Gamma(1+t)\bigl)\Bigl]$. As a consequence, 
the action for the bounce reduces to 

\begin{equation}\label{w2}
S_{\text{\tiny E}}(a(t),t)=\tau_{p}Z_{\text{\tiny E}}(a(t),t)
\quad.
\end{equation}

Before going on, it is time to remind the meaning of the entries on the right hand side of Eq.\eqref{w} 
as objects of the underlying worldsheet theory \cite{w}. Consider open bosonic string in Euclidean target space 
in the presence of  the tachyon background whose profile 
is similar to $T_{p+1}$ \footnote{Note that at the beginning this assumes the use of  parameters (bare couplings) 
which differ from the parameters in $T_{p+1}$. The 
parameters $a$ and $t$ are the renormalized couplings.}. This is a  simple choice of the background for which the 
partition function can be computed exactly. In a special scheme, it is given by $Z_{\text{\tiny E}}(a,t)$ \cite{w}. As 
to $\beta_a$ and $\beta_t$, they are the renormalization group (RG) beta functions. A simple RG analysis shows that 
the parameters flow from zero in the UV to infinity in the IR. The last means that all $X^i$ in the path 
integral are subject to the Neumann boundary conditions at the UV and the Dirichlet boundary conditions at the IR.

Having reminded the worldsheet theory, we have all at our disposal to achieve the purpose. First, let us note 
that the formula \eqref{w2} is the desired representation for the effective action \footnote{This relation 
is obvious for the UV as it directly follows from Eq.\eqref{w} but it is far from obvious for the IR.}. At this point, we have 
only to check that it results in a finite action for the bounce. $t$ is easily found from the correspondence 
between the fixed points of RG on the worldsheet and extrema of the effective action.  It is unique and given by its value in 
the IR fixed point. Next, plugging $t=\infty$ into the partition function, we get $Z_{\text{\tiny E}}(a(\infty),\infty)
\equiv Z_D=\bigl(4\pi^2\ap\bigr)^{\frac{p+1}{2}}$. From the worldsheet point of view, it is of course the expected result as 
the string path integral is finite for the Dirichlet boundary conditions. Second, the desired relation between the bounce 
and the D-instanton follows from a canonical construction of the D-instanton within the worldsheet theory (see, 
e.g. \cite{dj-rev}) as at $t=\infty$ all $X^i$ are subject to the Dirichlet boundary conditions. Finally, we finds for 
the exponential factor in the decay probability  

\begin{equation}\label{prob1}
w\sim\exp{\bigl[-\tau_{p} Z_D\bigr]}
\quad.
\end{equation}

Let us conclude by several short remarks:
\newline (i) So far, there is not known any partition function representation for string theory effective action that 
includes all string modes. It is the reason why we dealt with the background independent open string field theory. 
\newline (ii) Our analysis of the actions for the bounce is in fact similar to the computation of the descent relations 
between brane tensions in \cite{z, kud}.
\newline (iii) The idea that the D-instantons lead to nonperturbative effects like $\exp(-O(1/g))$ is an old 
one (see, e.g., \cite{pm,dj}). In particular, from the point of view \cite{dj}, the right hand side of Eq.\eqref{prob1} with 
$p=25$ is interpreted as the open string partition function with the Dirichlet boundary conditions for all $X^i$. 
This means that what we found can be called as the partition function representation for the exponential factor 
in a decay probability.
\newline (iv) We have made no attempt to study quantum corrections. It is clear that it would include higher 
genera of two-dimensional surfaces and, as a consequence, appearance of closed string modes. This makes 
the problem more involved than even including gravity within the field theory analysis \cite{cdl}. 
\newline (v) A perturbative instability of the standard bosonic open string vacuum ($T=0$ of Fig.1) has been 
discussed in \cite{1-loop}, where its decay rate has been evaluated via one-loop computations.

{\it Acknowledgements.} 
We have been benefited from discussions with L. Alvarez-Gaume, I. Bars, and J. Polchinski. We also would like to 
thank H. Dorn and A.A. Tseytlin for comments and reading the manuscript. This  work  is supported in part by DFG 
under grant No. DO 447/3-1 and the European Commission RTN Programme HPRN-CT-2000-00131. 


\end{document}